\documentclass{aastex631}
\usepackage{soulutf8}
\usepackage{ulem}

\def\ltsima{$\; \buildrel < \over \sim \;$}
\def\gtsima{$\; \buildrel > \over \sim \;$}
\def\lsim{\lower.5ex\hbox{\ltsima}}
\def\gsim{\lower.5ex\hbox{\gtsima}}

\shorttitle{New parameters to trace the dynamical evolution of star clusters}
\shortauthors{Bhat et al.}
\graphicspath{{./}{figures/}}

\begin{document}

\title{Searching for new observational signatures of the dynamical evolution of star clusters}

\author{B. Bhat}
\affiliation{Dept. of Physics and Astronomy "A. Righi", University of Bologna,
              Via Gobetti 93/2, Bologna, Italy}

\affiliation{INAF Osservatorio di Astrofisica e Scienza dello Spazio di Bologna, Via Gobetti 93/3, Bologna, Italy}

\author{B. Lanzoni}
\affiliation{Dept. of Physics and Astronomy "A. Righi", University of Bologna,
              Via Gobetti 93/2, Bologna, Italy}
\affiliation{INAF Osservatorio di Astrofisica e Scienza dello Spazio di Bologna, Via Gobetti 93/3, Bologna, Italy}

\author{F. R. Ferraro}
\affiliation{Dept. of Physics and Astronomy "A. Righi", University of Bologna,
              Via Gobetti 93/2, Bologna, Italy}
\affiliation{INAF Osservatorio di Astrofisica e Scienza dello Spazio di Bologna, Via Gobetti 93/3, Bologna, Italy}

\author{E. Vesperini}
\affiliation{Dept. of Astronomy, Indiana University, Bloomington, IN 47401, USA}

\begin{abstract}
We present a numerical study, based on Monte Carlo simulations, aimed
at defining new empirical parameters measurable from observations and
able to trace the different phases of star cluster dynamical
evolution.  As expected, a central density cusp, deviating from the
King model profile, develops during the core collapse (CC)
event. Although the slope varies during the post-CC oscillations, the
cusp remains a stable feature characterizing the central portion of
the density profile in all post-CC stages.  We then investigate the
normalized cumulative radial distribution (nCRD) drawn by all the
cluster stars included within one half the tridimensional half-mass
radius ($R \le 0.5 r_h$), finding that its morphology varies in time
according to the cluster’s dynamical stage.  To quantify these changes
we defined three parameters: $A_5$, the area subtended by the nCRD
within 5\% of the half-mass radius, $P_5$, the value of the nCRD
measured at the same distance, and $S_{2.5}$, the slope of the
straight line tangent to the nCRD measured at $R= 2.5\% r_h$.  The
three parameters evolve similarly during the cluster’s dynamical
evolution: after an early phase in which they are essentially
constant, their values rapidly increase, reaching their maximum at the
CC epoch and slightly decreasing in the post-CC phase, when their
average value remains significantly larger than the initial one, in
spite of some fluctuations. The results presented in the paper suggest
that these three observable parameters are very promising empirical
tools to identify the star cluster’s dynamical stage from
observational data.
\end{abstract}


\section{Introduction}
\label{sec:intro}
Globular clusters (GCs) are collisional stellar systems, where
frequent gravitational interactions among stars significantly alter
the overall energy budget, leading to a progressive internal change of
the cluster structure, and the radial distribution and content of
objects with different masses: the most massive stars tend to transfer
kinetic energy to lower mass objects (energy equipartition) and
progressively sink toward the system center (dynamical friction and
mass segregation); the energy transfer between stars causes the
gradual loss (evaporation) of stars (preferentially low-mass). The
continuous kinetic energy transfer from the core to the outskirts
leads to a runaway contraction of the core itself, with a substantial
increase of its density virtually toward infinity: the so-called
``core-collapse'' (hereafter, CC; see, e.g., \citealp{spitzer87,
  meylan+97}). The contraction is thought to be halted by the
formation and hardening of binary systems, and the post-CC phase is
characterized by several episodes of central density increase,
followed by stages during which the cluster rebounds toward a
structure with lower density and more extended core (the so-called
``gravothermal oscillations''; e.g., \citealt{meylan+97}).  Clearly,
these processes significantly affect the structure of the system with
respect to the initial conditions. Overall, the long-term internal
dynamical evolution tends to generate compact clusters, making
large-core systems naturally evolve toward objects with progressively
smaller core radius ($R_c$). Concurrently, the radial distribution of
stars with different masses progressively varies in time (the most
massive objects migrating to the center), and the high-density cluster
environment may facilitate the formation of exotic species that are
not predicted by the stellar evolution theory (such as blue straggler
stars, millisecond pulsars, low-mass X-ray binaries, intermediate-mass
black holes; e.g., \citealt{ferraro+97,ransom+05,pooley+03}).

The characteristic timescales of these changes depend in a very
complex way on the various internal and external properties, like
total cluster mass, initial size, central density, binary fraction,
orbit within the Galactic potential well, and so on. Hence, they can
significantly differ even in clusters of the same chronological
(stellar) age and, within the same system, from high- to low-density
regions.  Because of such a complexity, the observational
identification of the evolutionary stage reached by a cluster (i.e.,
its ``dynamical age'') may be hard and lead to ambiguous conclusions.
Of course, this may significantly hamper efforts aimed at linking the
theoretical predictions concerning the dynamics of star clusters with
observations.  In this respect, due to their proximity to the Sun,
Galactic GCs (GGCs), represent the ideal laboratory (possibly the only
place in the Universe) where dynamical evolutionary processes can be
investigated in thorough detail. The vast majority of these stellar
systems is very old (t=12 Gyr, comparable to the age of the Universe;
see, e.g., \citealt{marinfranch+09, forbes+10}), but due to different
internal and external properties, they sample essentially all the
dynamical evolutionary stages expected for multi-body and multi-mass
systems.

Up to recently, the characterization of the dynamical age of GGCs was
essentially based on their structural morphology. In fact, one of the
most used indicators, the central relaxation time ($t_{\rm rc}$), is
estimated from the measure of structural parameters such as the core
radius and the central density, following the analytical expression
proposed, e.g., in eq.(10) of \citet[][see also \citealp{spitzer87}]{djo93}.  
These parameters are usually ``read''
from the King model \citep{king66} that best reproduces the projected
density profile of a star cluster.  Also the classification of GGCs as
CC or post-CC systems has been so far based on the detection of a
morphological feature, i.e., a steep power-law cusp in the central
portion of the density profile (see \citealt{djo+84, ferraro+03,
  ferraro+09}), strongly deviating from the flat-core behavior of the
King model, which well reproduces the density distribution of non-CC
clusters.  This diagnostic, however, may be not fully reliable since
the cusp may be significantly reduced during the post-CC gravothermal
oscillations, or due to the occurrence of other processes (e.g.,
depending on the binary fraction) that may contribute to significantly
delay/reduce the ``intensity'' of CC, or it can be hardly detectable
from observations. As a matter of fact, only a small fraction
(15-20\%) of the entire population of GGCs displays a central cusp in
the star density profile and is classified as post-CC (see
\citealt{djo+84,lugger+95}), in spite of the fact that the central
relaxation time is sensibly shorter than the age in most of the cases
(see the compilations by \citealt{djo93} and \citealt{harris96}).

GGCs also provide the advantage that stars can be resolved and studied
individually, thus offering additional possibilities to investigate
the internal dynamical state of the system.  In principle, either
specific classes of objects, or the entire cluster population can be
used as probes of the cluster dynamical evolution.  Indeed, a lot of
work in this direction has been done over the last decades.  Several
theoretical works, mainly based on the results of extensive N-body
simulations, recently suggested that radial variations of the stellar
mass function, the presence of orbital anisotropy, and the velocity
dispersion profile as a function of stellar mass can be used to infer
the level of energy equipartition and the dynamical state of GCs
\citep[e.g.,][] {baumgardt_makino03, tiongco+2016, bianchini+2016,
  webb_vesperini2017, bianchini+2018}. In turn, these diagnostics are
becoming measurable in an increasing number of GGCs, especially thanks
to multi-epoch HST observations and improved procedures of data
analysis, which allow high-precision photometry and proper motion
measurements for stars down to a few magnitudes below the main
sequence turn-off (MS-TO). Recent examples of this kind of studies can
be found, e.g., in \citet{libralato+2018, libralato+2019}, and
\citet{cohen+2021}. While these approaches require observations that,
even in the Gaia era, are still very challenging for most GGCs (due to
their high central densities and relatively large distances from
Earth), our group focused the attention on a special class of exotic
objects, the so-called blue straggler stars (BSSs), which possibly
offer the most promising and viable way to trace the dynamical
evolution of dense stellar systems.  Being generated by direct
collisions \citep{hillsday, lombardi95, sills05} or mass-transfer
activity in binaries \citep{mccrea, leonard96}, they turn out to be
significantly heavier ($M_{\rm BSS}=1.2 M_\odot$; \citealp{ferraro+06,
  lanzoni+07a, fiorentino+14, raso+19}) than the average cluster
population ($\langle m\rangle =0.3 M_\odot$). Hence, these stars are
powerful gravitational probes of key physical processes (such as mass
segregation and dynamical friction) characterizing the dynamical
evolution of star clusters. In fact, the radial distribution of BSSs
has been used as ``dynamical clock'' to efficiently measure the
dynamical aging of stellar systems \citep{ferraro+12, lanzoni+16,
  ferraro+20}: the level of BSS central segregation with respect to
normal (lighter) stars allowed the ranking of GGCs in terms of their
dynamical age, from dynamically young systems (with negligible BSS
segregation), to highly dynamically evolved clusters, where BSSs are
much more centrally concentrated than the reference population
\citep{ferraro+12, ferraro+18, lanzoni+16}. The level of BSS central
segregation is measured via the $A^+$ parameter, defined as the area
between the cumulative radial distribution of BSSs and that of a
lighter, reference population \citep{alessandrini+16}. A strong
correlation between $A^+$ and the number of relaxation times occurred
since cluster formation has been found from the analysis of $\sim 1/3$
of the entire GGC population \citep{ferraro+18} and it is confirmed
also for a sample of old GCs in the Large Magellanic Cloud
\citep{ferraro+19}.

BSSs have been found to also trace the occurrence of CC and probe the
time when it happened.  In fact, the double BSS sequence detected for
the first time in the post-CC cluster M30 \citep{ferraro+09} has been
interpreted as the manifestation of the two formation processes, with
the bluest sequence being populated by collisional BSSs generated by
an enhanced activity of gravitational interactions during
CC. Moreover, the measure of the extension of the blue sequence
provided the first empirical dating of the CC event (see
\citealt{ferraro+09, portegies19}). Since then, the double BSS
sequence has been discovered in several additional post-CC clusters
(see the cases of NGC 362 in \citealt{dalessandro+13}, M15 in
\citealt{beccari+19}, and NGC 6256 in Cadelano et al. 2021, in
preparation), thus strengthening the link between this feature and the
CC event.

An obvious limitation of using BSSs to infer the dynamical state of
the host stellar system is that they are few in number (of the order
of a few dozen, on average). Moreover, having different diagnostics of
internal dynamical evolution is certainly desirable and useful. Hence,
here we analyze the time evolution of a ``synthetic GC'' obtained from
a Monte Carlo simulation run, with the specific aim of defining
suitable diagnostics of dynamical aging from a new perspective, i.e.,
by using the entire population of evolved stars. We provide the
definition of three new parameters and test their effectiveness in
distinguishing clusters in the pre-CC phase, from those experiencing
post-CC evolutionary stages, thus tracing the dynamical aging of the
system up to CC and beyond. The paper is organized as follows. In
Section \ref{sec:simu} we describe the initial conditions of the Monte
Carlo simulation run, and the (observational) approach adopted in the
following analysis.  Section \ref{sec:dens} discusses the method used
to determine the projected density profile and the best-fit
\citet{king66} model of each extracted snapshot. In Section
\ref{sec:nCRD} we present the assumptions adopted to build the
normalized cumulative radial distributions of cluster stars and
discuss the dependency of their morphology on the simulation
evolutionary time.  This is then used in Section \ref{sec:params} to
define three new empirical parameters able to trace the internal
dynamical evolution of stellar systems. Their dependency on the
adopted assumptions is discussed in Section \ref{sec:sensitivity}. The
summary and conclusions of the work are presented in Section
\ref{sec:discussion}.

\section{Initial Conditions and Methods}
\label{sec:simu}
In this work we focus our attention on the dynamical evolution of a
star cluster followed with a Monte Carlo simulation run with the MOCCA
code (\citealt{Hypki_Giersz_2013,Mocca_giersz}). The code includes the
effects of binary and stellar evolution (modeled with the SSE and BSE
codes; \citealt{Hurley_2000,Hurley_2002}) with supernovae kicks
assumed to follow a Maxwellian distribution with dispersion equal to
265 km/s \citep{Hobbs_05}, the effects of two-body relaxation and a
tidal truncation. The initial conditions of the simulation have been
chosen well within the range of values observed for GGCs, with the
main aim to provide us with a system that experiences all dynamical
evolutionary phases and reaches CC within 12-13 Gyr from
formation. Since the prime goal of the paper is to put forward the
definition of new dynamical indicators, in the following we will
present the detailed analysis of this specific run. However, two
additional simulations, run from slightly different initial
conditions, are discussed in Appendix A, and a much more extensive
exploration of the parameter space and its effects on the final
results will be the subject of forthcoming dedicated studies (see
Section \ref{sec:discussion} for more details).  The simulated cluster
has initially 500K single stars with masses ranging between $0.1
M_\odot$ and $100 M_\odot$ following a \citet{IMF} mass function.  The
initial total mass of the system is $\sim 3.2 \times 10^5 M_\odot$,
while it is approximately half this value after 12 Gyr of evolution.
The stars are initially distributed as a \citet{king66} model with
dimensionless central potential $W_0 = 6$, and the cluster is tidally
underfilling, with a tridimensional half-mass radius $r_h = 2$ pc and
a Jacobi radius set equal to 61 pc (corresponding to the value the
cluster would initially have if orbiting at a Galactocentric distance
equal to 4 kpc). No primordial binaries are included in this run,
although binary stars dynamically form as the system approaches the CC
phase. Since the code includes prescriptions for stellar evolution, it
provides for every star at any evolutionary time not only the mass and
the three components of position and velocity, but also the magnitude
in two photometric bands (namely the V and B band), from which a
color-magnitude diagram (CMD) of the stellar content at any epoch can
be built.

The simulation follows the cluster evolution for $\sim 16$ Gyr from
its formation. Although this is larger than the Hubble time, it allows
us to follow the cluster dynamical evolution also after CC, which
occurrs at 12.8 Gyr (see below). In Figure \ref{fig:lagrange_rad} we
show the time evolution of the cluster's 1\% Lagrangian radius
($r_{1\%}$, i.e., the radius including 1\% the total cluster's mass):
the temporal variation of this radius illustrates well the various
phases of the cluster's dynamical evolution and, in particular, the CC
and post-CC phases. The effects of mass loss due to stellar evolution
cause the cluster to initially expand as shown by the early increase
in $r_{1\%}$. Then, after $\sim 2$ Gyr from formation, two-body
relaxation starts to drive the evolution of the cluster central region
leading it to a progressive contraction, with $r_{1\%}$ shrinking by
approximately a factor of 7 in 10 Gyrs.  The figure shows that the
contraction phase is characterized by an initial ``slow phase''
($r_{1\%}$ shrinks by a factor of 2 in approximately 8 Gyrs) and a
final ``rapid phase'' (a factor 4 shrinking, from 0.25 pc to 0.07 pc,
in less than 3 Gyrs). At this time, $r_{1\%}$ reaches its minimum
value: this is the CC event, occurring at a time $t_{\rm CC}=12.8$
Gyr. Later, a phase characterized by gravothermal oscillations is
clearly distinguishable in the figure, as cyclic expansions and
contractions of the $1\%$ Lagrangian radius. Of course, during this
evolution, not only the central region, but the entire cluster
structure varies with time. To carefully investigate these changes, we
extracted 38 time snapshots sampling different evolutionary phases of
the system, as marked by the vertical dotted lines in Figure
\ref{fig:lagrange_rad}. In order to easily and immediately link each
snapshot to the corresponding evolutionary phase, we adopted the
following color code: {\it green} for snapshots belonging to the early
slow contraction phase, {\it cyan} for snapshots belonging to the
final rapid contraction phase, {\it blue} for snapshots sampling the
CC phase, and {\it yellow} for snapshots probing the post-CC
gravothermal oscillations epoch.

\begin{figure}
\centering
\includegraphics[scale=0.7]{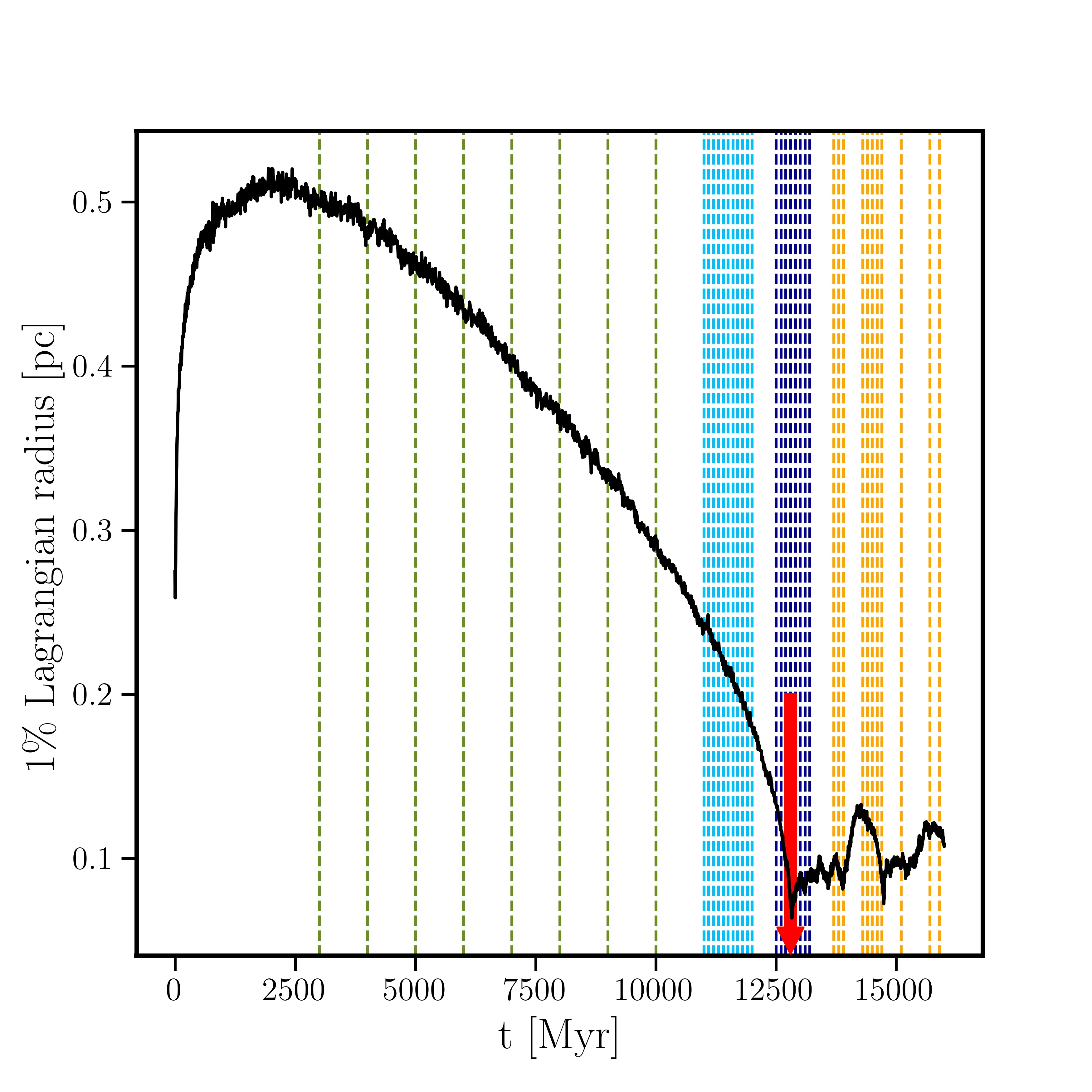}
\caption{Time evolution of the 1\% Lagrangian radius (in pc) of the
  simulated cluster (black line). The vertical lines correspond to the
  38 time snapshots analyzed in this work, color-coded as follows to
  mark different evolutionary stages: green, cyan, blue and yellow for
  early, pre-CC, CC, and post-CC, respectively. The time of CC
  ($t_{\rm CC}=12.8$ Gyr) is marked with a large red arrow.}
\label{fig:lagrange_rad}
\end{figure}

In the following analysis, each time snapshot has been studied from an
``observational perspective'', i.e., as if the simulation output was
the product of real observations.  This is meant to allow the
definition of parameters that can be realistically derived and
measured in observational investigations. Thus, procedures and
standard strategies, as well as approximations routinely adopted in
dealing with observational data, were applied to the snapshots. To
this purpose, each snapshot has been projected on a 2D plane and the
distances of all the stars from the cluster's center have been
transformed from parsecs to arcseconds assuming that the system is at
10 kpc from the Sun, which is the typical distance of GGCs
\citep{harris96, baumgardt_vasiliev2021}. We also limited most of the
analysis only to stars that are brighter than 0.5-1 magnitudes below
the MS-TO, in agreement with the threshold adopted in many
observational studies to avoid photometric incompleteness biases and
to deal with samples of equal-mass stars \citep[see,
  e.g.][]{lanzoni+07b, lanzoni+10, lanzoni+19, miocchi+13}.

\section{RESULTS}
\label{sec:resu}
\subsection{Projected density profile}
\label{sec:dens}
\begin{figure}
\centering
\includegraphics[scale=0.35]{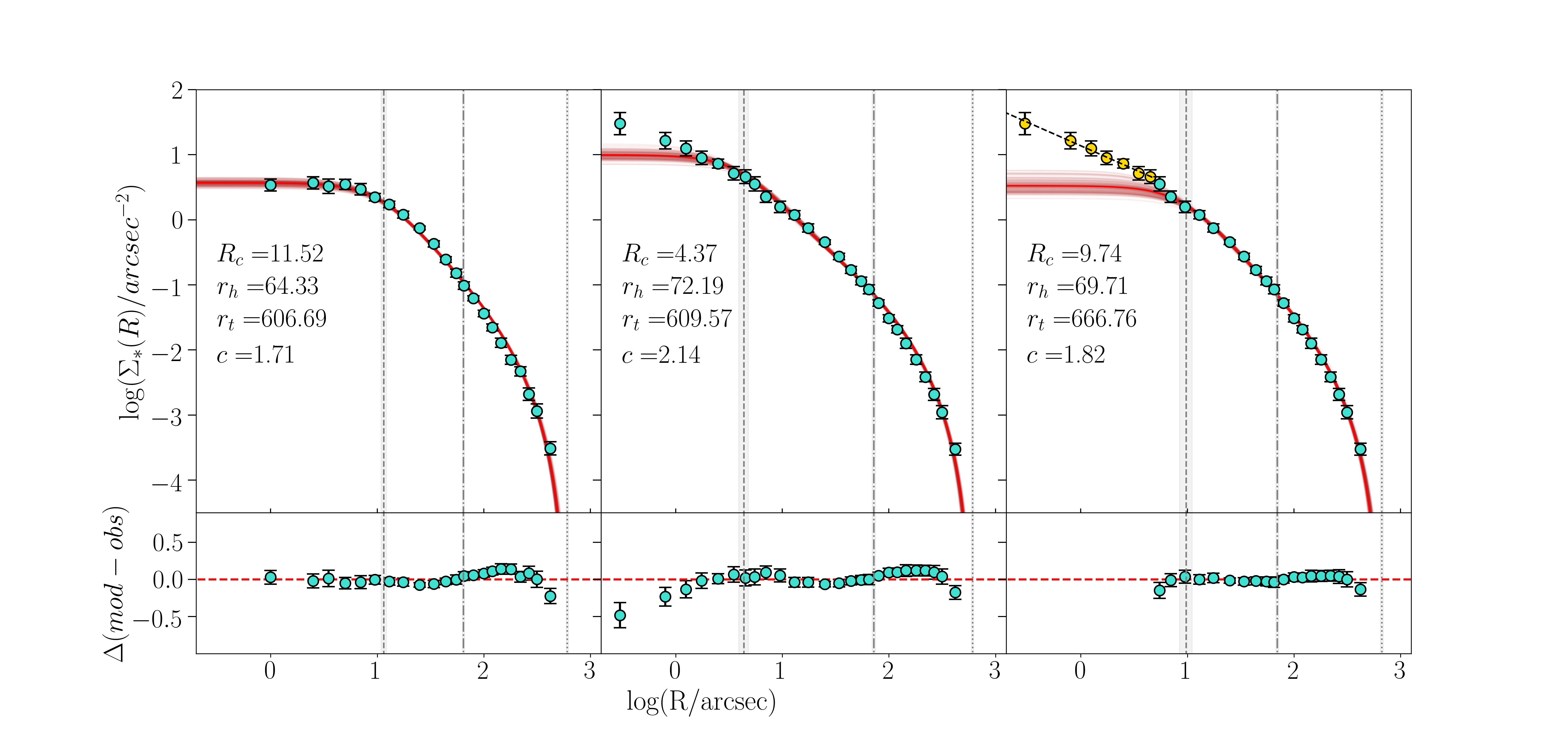}
\caption{Projected density profile (colored circles), obtained as
  number of stars per unit area in different concentric annuli around
  the cluster center, for two simulation snapshots: a pre-CC cluster
  at $t=7$ Gyr (left panel), and a post-CC system at $t=13.8$ Gyr
  (central and right panels).  The thick red line corresponds to the
  best-fit King model, while the shaded area marks its uncertainty
  (see Section \ref{sec:dens} for the details). The bottom panels show
  the residuals between the observations and the best-fit King
  model. For the post-CC snapshot, two different approaches have been
  adopted to determine the best-fit King solution: (1) the entire
  observed profile has been considered (central panel, cyan circles),
  (2) only the data points beyond $5\arcsec$ from the center have been
  included in the fit (right panel, cyan circles), while the innermost
  portion of the profile (yellow circles) has been described through a
  linear fit (black dashed line). The positions of the core, half-mass
  and tidal radii are marked, respectively, by a dashed,
  dotted-dashed, and dotted vertical line, and their values are
  labelled in the legend together with that of the concentration
  parameter.}
\label{fig:density_profile}
\end{figure}

As a first step of our analysis, we studied the projected density
profile of the simulated cluster at different epochs, to verify
whether a central cusp develops at $t_{\rm CC}$, as expected, and how
it evolves with time. To this purpose, we followed the same procedure
adopted in several observational works determining the radial
distribution of stellar number counts per unit area, $\Sigma_*(R)$,
instead of the surface brightness profile \citep[e.g.][]{miocchi+13,
  lanzoni+19}. Summarizing: {\it (i)} only stars brighter than one
magnitude below the MS-TO (i.e. with $V<V_{\rm TO}+1$) have been taken
into account; {\it (ii)} the sampled area has been divided into
concentric annuli centered on the cluster center, assumed to be at
coordinates (0,0), and {\it (iii)} each annulus has been typically
partitioned into four sub-sectors. The exact number of annuli and
sub-sectors is chosen as a compromise between including a sufficiently
large number of stars to provide enough statistics, and a good radial
sampling of the profile. Thus, it was set according to the
(time-evolving) structure of the system. The projected cluster density
at every radial distance from the center was then determined as the
average number density of particles in the adopted sub-sectors, and
its uncertainty was estimated from the variance among the sub-sectors.
For the sake of illustration, in Figure \ref{fig:density_profile} we
show the projected star density profile obtained for two
representative snapshots: one determined at $t=7$ Gyr, during the
pre-CC evolution (left panel), the second obtained for $t=13.8$ Gyr,
slightly after CC(middle and right panels).  As expected, the central
portion of the former is flat, while a significant density cusp,
following a steep power-law behavior, is clearly visible toward the
cluster center in the post-CC case.

Following what is commonly done in observational works
\citep[e.g.][]{miocchi+13, lanzoni+19}, we then searched for the
single-mass \citet{king66} model that best fits the density profile
obtained in the various snapshots. We explored a grid of models with
dimensionless parameter $W_0$ (which is proportional to the
gravitational potential at the center of the system) varying between 4
and 10.75 in steps of 0.05, corresponding to a concentration parameter
$c$ spanning the interval between 0.84 and 2.5. This parameter is
defined as $c=\log(r_t/r_0)$, where $r_t$ is the truncation or tidal
radius of the system, and $r_0$ is the characteristic scale-length of
the model named ``King radius".  The latter is often identified with
the core radius $R_c$, which is the observationally accessible scale
length corresponding to the distance from the center where the
projected density is equal to half the central value.  Indeed, they
are quite similar, especially for large values of $W_0$ or $c$: the
ratio $R_c/r_0$ varies between $\sim 0.82$ for $c=0.84$, and $\sim
0.99$ for $c=2.5$.  We adopted the $\chi^2$ approach described in
detail in \citet[][see also \citealp{miocchi+13}]{lanzoni+19} to
determine the best-fit solution (i.e., the one minimizing the
residuals between the model and the ``observed'' profile) and to
estimate the uncertainties of the best-fit parameters.

In agreement with what is found observationally for most GCs, we
conclude that the King model family well reproduces the projected
density profiles of pre-CC systems, while it shows a clear
inconsistency in the innermost region of CC and post-CC
snapshots. Figure \ref{fig:density_profile} illustrates the result for
the pre- and a post-CC cases discussed above.  The best-fit King model
is shown as a thick red line, while its uncertainty is represented by
the shaded area and corresponds to the set of King models built by
varying the fitting parameters within their uncertainty ranges.  The
King model function excellently reproduces the observed profile at any
distance from the cluster center for the pre-CC snapshot (left
panel). Conversely, being constant at small radii by construction, it
cannot properly describe the central density cusp observed at $t=13.8$
Gyr, after CC (central panel).  Hence, in the presence of a central
density cusp two different approaches are possible: either (1) to fit
the entire density profile and search for the model providing the best
solution regardless of its inadequacy in the region close to the
center, or (2) to exclude from the fit the innermost portion of the
density profile (here we assumed $R<5\arcsec$ as a reasonable value).
The central panel of Figure \ref{fig:density_profile} corresponds to
approach (1), while the right-hand panel shows the result of approach
(2). The latter, combined with a linear fit to the innermost data
points (oblique dashed line in the figure), clearly allows a much
better description of the complex shape of the star density profile of
CC and post-CC systems, but it depends on the (arbitrary) choice of
the size of the region to be excluded from the King fit and it
provides parameters (as $R_c$, $r_h$ and $c$) that are not
representative of the real cluster structure. Approach (1) clearly
provides a poor representation of the central density profile, but it
offers the advantage to be free from arbitrariness, thus allowing a
coherent analysis of the density profile irrespective of the cluster
dynamical stage.
  
\begin{figure}
\centering
\includegraphics[scale=0.5]{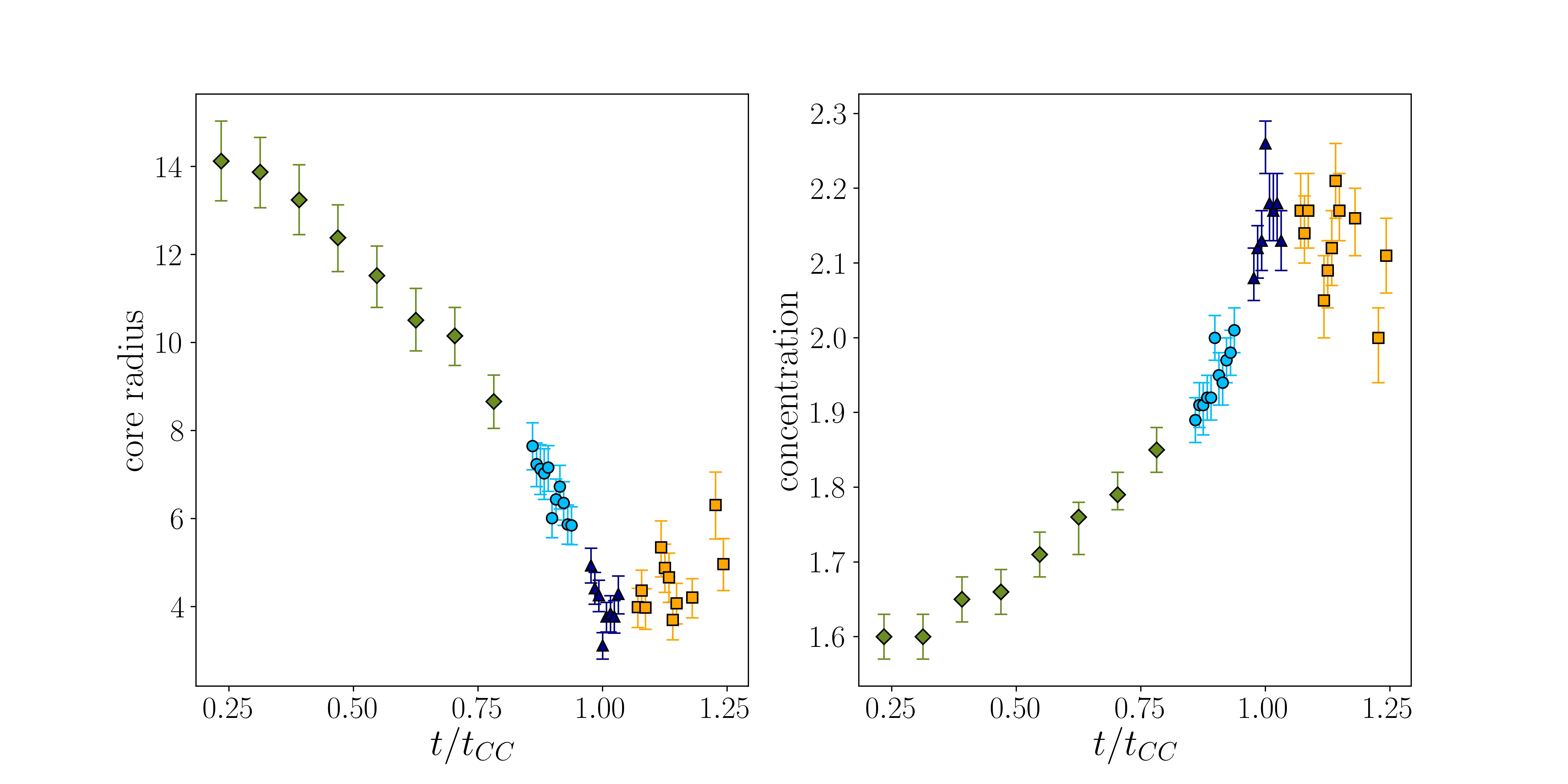}
\caption{Time evolution of the core radius (left panel) and
  concentration parameter (right panel) as obtained from the King
  model fit to the entire star density profile of all the extracted
  snapshots. Different colors and symbols mark different evolutionary
  stages: green diamonds, cyan circles, blue triangles and yellow
  squares for early, pre-CC, CC, and post-CC, respectively (the color
  code is the same as in Figure \ref{fig:lagrange_rad}).}
\label{fig:rc}
\end{figure}

Although defining a core region (with constant density) is formally
meaningless for CC and post-CC systems, the time behavior of $R_c$ and
$c$ obtained from approach (1) is qualitatively consistent with that
of the 1\% Lagrangian radius. This is illustrated in Figure
\ref{fig:rc}, showing that the core radius progressively decreases to
a minimum value at $t_{\rm CC}$, then stays almost constant for
increasing time, similarly to the trend of $r_{1\%}$ shown in Figure
\ref{fig:lagrange_rad}. The concentration parameter $c$ displays the
opposite behavior and when the power law cusp develops, during and
after CC, it reaches values larger than $\sim 2$. The time evolution
of these parameters is in agreement with previous findings
\citep[e.g.][]{simulation1, simulation2}. Indeed a concentration
parameter $c\sim 2$-2.5, together with the the presence of a central
cusp in the density profile, are the two diagnostics commonly used in
the literature to classify a GC as CC or post-CC (see, e.g.,
\citealt{djo+84, lugger+95, ferraro+09}).

\begin{figure}
\centering
\includegraphics[scale=0.4]{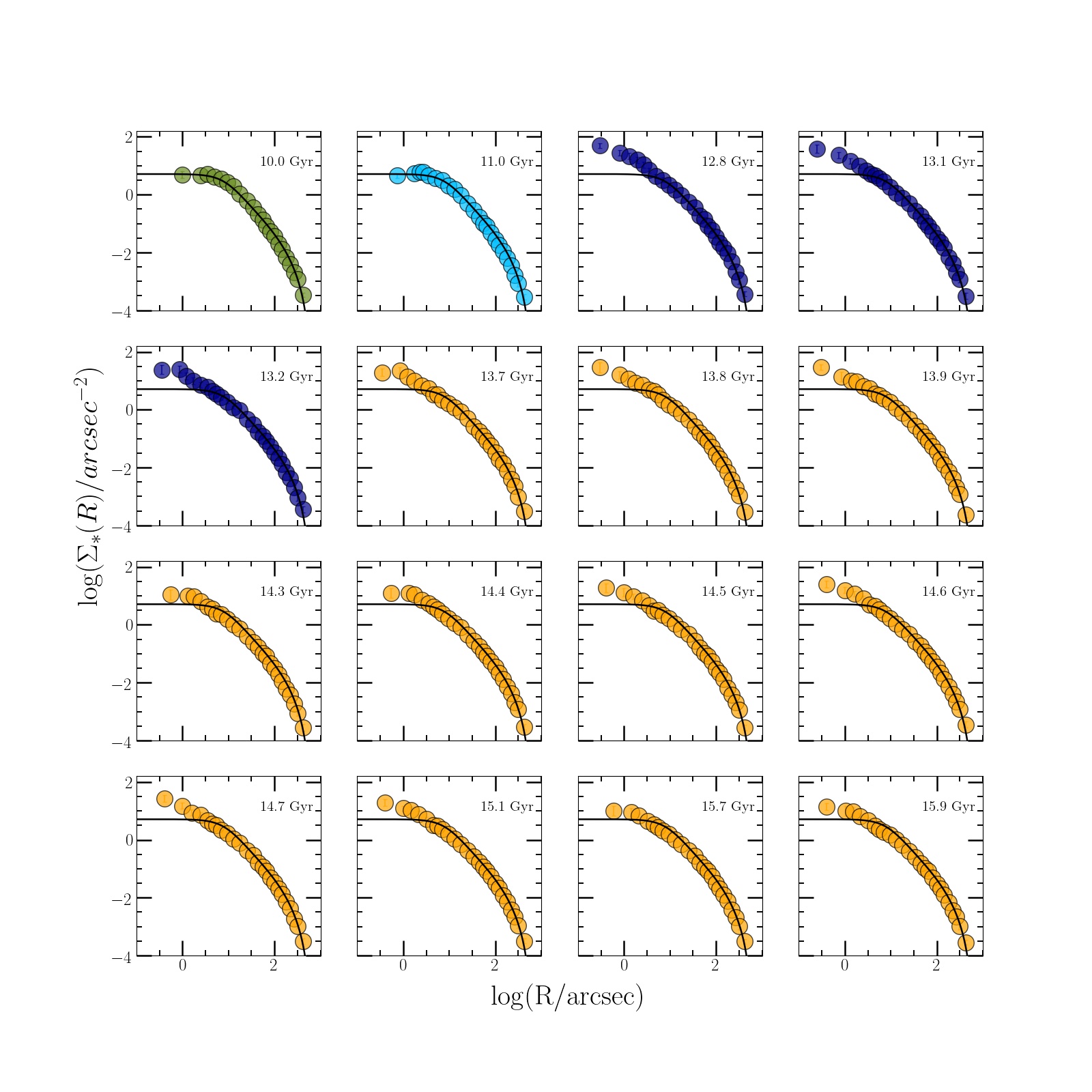}
\caption{Comparison among the projected star density profiles measured
  during the latest stages of dynamical evolution, illustrating the
  appearance and persistence of the central cusp.  All the profiles
  are normalized to the density measured at $R=65\arcsec$ in the 10
  Gyr snapshot, and the black solid line is the best-fit King model to
  the 10 Gyr density profile.  The color code is as in Figure
  \ref{fig:lagrange_rad}: cyan for pre-CC snapshots, blue for the CC
  phase, and yellow for post-CC stages and the snapshot time is
  labelled in each panel.}
\label{fig:cc}
\end{figure}

To properly explore the development and transformation of the central
cusp in the simulations, we show in Figure \ref{fig:cc} the star
density profile of the latest 16 snapshots, sampling the last 6 Gyr of
evolution (from 10 to 16 Gyr). To facilitate the comparison, all the
profiles have been vertically shifted until their density at
$R=65\arcsec$ matches the one measured in the $t=10$ Gyr snapshot,
with $65\arcsec$ ($\sim 3$ pc) roughly corresponding to the value of
the tridimensional half-mass radius of the 10 Gyr best-fit King model,
which is represented in all panels as a black solid line.  The color
code is the same adopted in Figure \ref{fig:lagrange_rad} to flag the
dynamical stage of the selected snapshots.  The first two panels show
that $\sim 3$ and 2 Gyr before CC, respectively, the star density
profile is well reproduced by the King model. The third panel samples
the CC event and the setting of the cusp. The remaining 13 panels
probe the gravothermal oscillation phase. As can be seen, once set,
the central cusp remains visible in the whole post-CC evolution
(yellow profiles). This is the first relevant result of the present
analysis, since it clearly demonstrates that the central cusp, once
set, never disappears, in spite of the subsequent core radius
oscillations. Hence, we can conclude that the density profile of star
clusters is characterized by the presence of a central cusp also
during the post-CC gravothermal oscillation phase.

However, the simulation also shows that the cusp's slope varies in
time during the post-CC stage and the cusp becomes shallower during
the expansion phases of gravothermal oscillations, thus rising the
problem of its operational detectability and proper characterization.
Indeed, the cusp detection and its correct measure are among the most
critical issues from the observational point of view. The photometric
incompleteness of the catalog, which becomes increasingly severe in
the innermost cluster regions, artificially decreases the number of
resolved stars close to the center. Thus, an appropriate assessment of
the level of completeness of the observational sample is a key step to
firmly establish the existence and the entity of the cusp.  If the
density profile is built with methods similar to that described above,
but too shallow observations are used, the resulting low statistics
may force the use of too large radial bins, which directly affects the
ability of detecting the cusp.  Also the exact definition of the
innermost cluster region where the cusp manifests itself can affect
the significance of the deviation from a King (centrally flat) model.
This region is not known \emph{a priori} and different assumptions
about its extension might lead to either an over-, or under-estimate
of the cusp steepness.  All this raises the need for new indicators of
the internal dynamical state of dense stellar systems, not depending
on the detection of a central density cusp or lack thereof.

\subsection{The Normalized Cumulative Radial Distribution}
\label{sec:nCRD}
Similar to the route followed to refine the definition of the
``dynamical clock'' based on BSSs (compare, e.g., \citealp{ferraro+12}
with \citealp{lanzoni+16} and \citealp{ferraro+18}), here we explore a
new way to infer the dynamical evolutionary stage of a GC using the
normalized cumulative radial distribution (nCRDs) of its stellar
population. In particular, for every simulation snapshot, we consider
all the stars brighter than a threshold $V_{\rm cut}$ located within a
projected distance $R_n$ from the center, and we determine their
nCRD. By construction, this function varies between 0 (at $R=0$) and 1
(at $R=R_n$), describing, for each value of $R$, the percentage of
stars within that distance from the center (i.e., the number of stars
counted within $R$ normalized to the total number of stars within
$R_n$). The magnitude cut has the purpose of mimicking the analysis of
observed data sets, where it is needed to avoid photometric
incompleteness biases and/or is set by the exposure time of the
available images. Consistently with many observational studies, we
adopted $V_{\rm cut}=V_{\rm TO}+0.5$.  The choice of a normalization
radius ($R_n$) has the aim to refer the analysis to the same
\emph{physical} region in all snapshots, thus allowing a direct
comparative evaluation of the effects of dynamical evolution in
clusters of different sizes and in different dynamical stages.  We
built the nCRDs for several values of $R_n$, concluding that $R_n =
0.5\times r_h$ is the best choice, because it maximizes the
morphological differences that dynamical evolution imprints on the
nCRD (see below), while still providing large statistics. Indeed,
$0.5\times r_h$ is a distance from the center small enough to be
highly sensitive to dynamical evolutionary effects (which are
strongest in the most central regions) and large enough to include
large samples of stars. We emphasize that $r_h$ is defined as the
(tridimensional) radius of the sphere that includes half the total
cluster mass, while $R_n = 0.5\times r_h$ is the projected distance
from the center within which selecting the stars to build the
nCRD. While $r_h$ is not directly observable, its value is
unambiguously obtained from the King model that, in projection,
best-fits the observed density profile.  To avoid the arbitrariness of
approach (2) to the fit of the star density profile (see discussion in
Section \ref{sec:dens}), for all the snapshots we adopted the value of
$r_h$ obtained from the King model that best reproduces the entire
density distribution (see an example in the central panel of Figure
\ref{fig:density_profile}).

\begin{figure}
\centering
\includegraphics[trim={0 0 0 2.1cm},clip,width=150mm,scale=0.9]{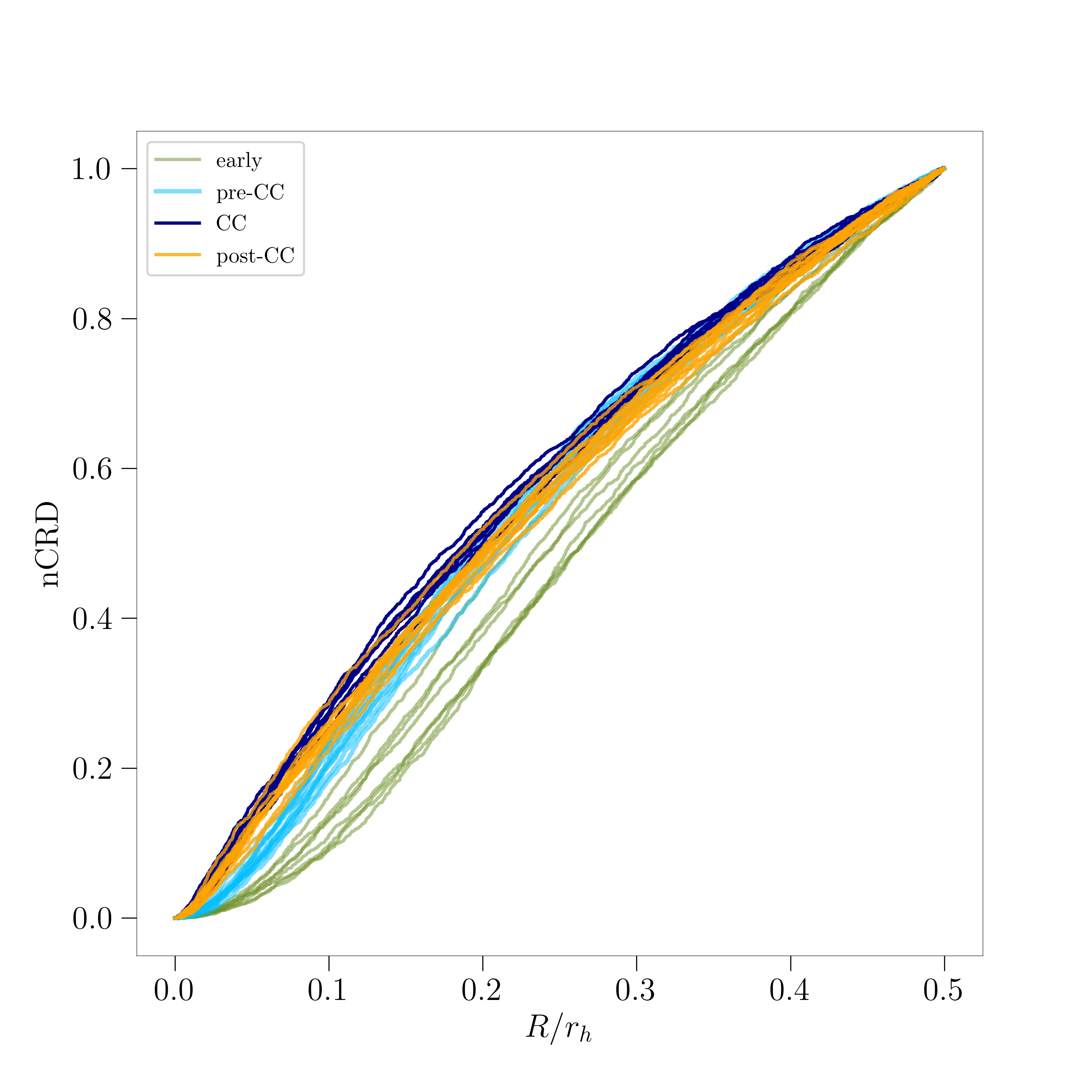}
\caption{Normalized cumulative radial distributions of all the stars
  with $V=V_{\rm TO}+0.5$ and $R< 0.5 \times r_h$, for all the
  analysed simulation snapshots. The nCRDs are plotted according to
  the color code adopted in Fig. \ref{fig:lagrange_rad}: from early
  times (green), to pre-CC stages (cyan), the CC phase (blue), and the
  post-CC gravothermal oscillation epoch (yellow).  }
\label{fig:CRD}
\end{figure}

Figure \ref{fig:CRD} shows the nCRDs of all the considered snapshots,
plotted with different colors according to the dynamical state, as in
Fig.\ref{fig:lagrange_rad}.  As can be seen, the nCRDs differ one from
the other and they do not appear to be randomly arranged: a nice
progression of color groups following the aging sequence defined in
Fig.\ref{fig:lagrange_rad} is clearly visible, from green (early
stages), to cyan, blue, and yellow (late, post-CC stages). This
indicates that the nCRD of the cluster population is sensitive to the
parent cluster dynamical age.  As a consequence, the dynamical stage
of a GC should be measurable from an appropriate parametrization of
the morphology of its nCRD.

\subsection{Defining the new parameters}
\label{sec:params}
\begin{figure}
\centering
\includegraphics[scale=0.5]{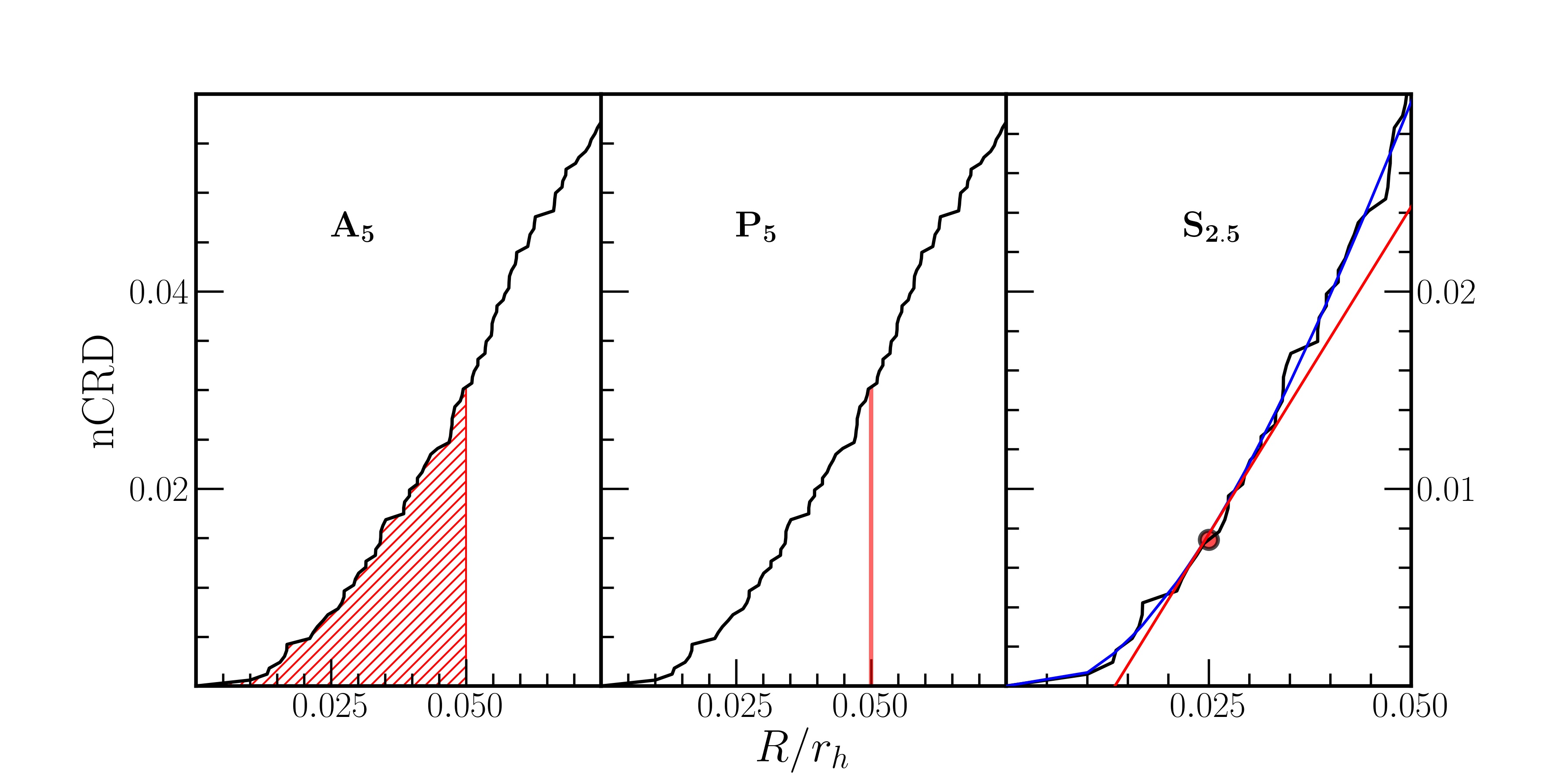}
\caption{Definition of the three new diagnostics of dynamical
  evolution based on the nCRDs of the cluster stellar population shown
  in Figure \ref{fig:CRD}. $A_5$ is the area subtended by the nCRD
  within a projected distance equal to 5\% the half-mass radius (red
  shaded region in the left panel).  $P_5$ is the percentage of stars
  measured at $R=0.05\times r_h$, corresponding to the value of the
  nCRD at this clustercentric distance. $S_{2.5}$ is defined as the
  slope of the straight line tangent to the nCRD at $R=2.5\%$ the
  half-mass radius, as illustrated in the right panel: operationally,
  it is measured from the tangent line (red line) at $R =0.025\times
  r_h$ (large red circle) to the polynomial function that best
  reproduces the nCRD (blue line).}
\label{fig:para}
\end{figure}

The result shown in Figure \ref{fig:CRD} clearly shows that the
cluster dynamical aging is imprinted in the morphology of the nCRD of
its stellar population. Both the percentage of stars within a given
distance from the cluster center and the growth rate of the nCRD as a
function of the clustercentric distance appear significantly different
at different stages of dynamical evolution. Hence, both these
quantities, in principle, could be used to quantify the dynamical
state of a GC. Not surprisingly, the differences among the nCRDs appear
more pronounced in the innermost radial portion of the system, where
the effects of dynamical evolution are known to be stronger. We thus
defined the following three parameters to quantify the nCRD
differences:
\begin{itemize}  
\item[1)]{\bf $A_5$ -- } It is defined as the area subtended by each
  nCRD between the center ($R=0$) and a projected distance equal to
  5\% the tridimensional half-mass radius ($R=0.05\times r_h$), as
  illustrated by the shaded region in the left-hand panel of Figure
  \ref{fig:para}.  Because of the progressive increase of the central
  density during dynamical evolution, $A_5$ is expected to increase
  with time.
\item[2)] {\bf $P_5$ -- } It is defined as the value at $R=0.05\times
  r_h$ of the nCRD defined as above and it is illustrated in the
  central panel of Figure \ref{fig:para}. The progressive contraction
  of the system toward CC translates into a centrally steeper nCRD
  and, as a consequence, also the value of this parameter is expected
  to increase as a function of the cluster dynamical age.
\item[3)]{\bf $S_{2.5}$ -- } It is defined as the slope of the
  straight line tangent to the nCRD at a projected distance equal to
  2.5\% the half-mass radius ($R=0.025\times r_h$).  Operationally, a
  polynomial function (in the form $y=a\times x^3+b\times x^2+ c\times
  x$, with $x=R/r_h$) is fitted to the nCRD to smooth out its noisy
  behavior, and $S_{2.5}$ is the slope of the straight line tangent to
  the polynomial (see the red and the blue lines, respectively, in the
  right panel of Figure \ref{fig:para}).  Since our analysis showed
  that the most relevant changes in the growth rate of the nCRD occur
  in the very internal region of the system, $S_{2.5}$ has been
  defined at an even smaller projected distance from the center with
  respect to the other two parameters (namely, at 2.5\%, instead of
  5\%, the half-mass radius).  It quantifies the radial growth rate of
  the nCRD and, similarly to $A_5$ and $P_5$, it is expected to
  increase as function of the cluster dynamical age.
\end{itemize}    

\begin{figure}
\centering
\includegraphics[scale=0.5]{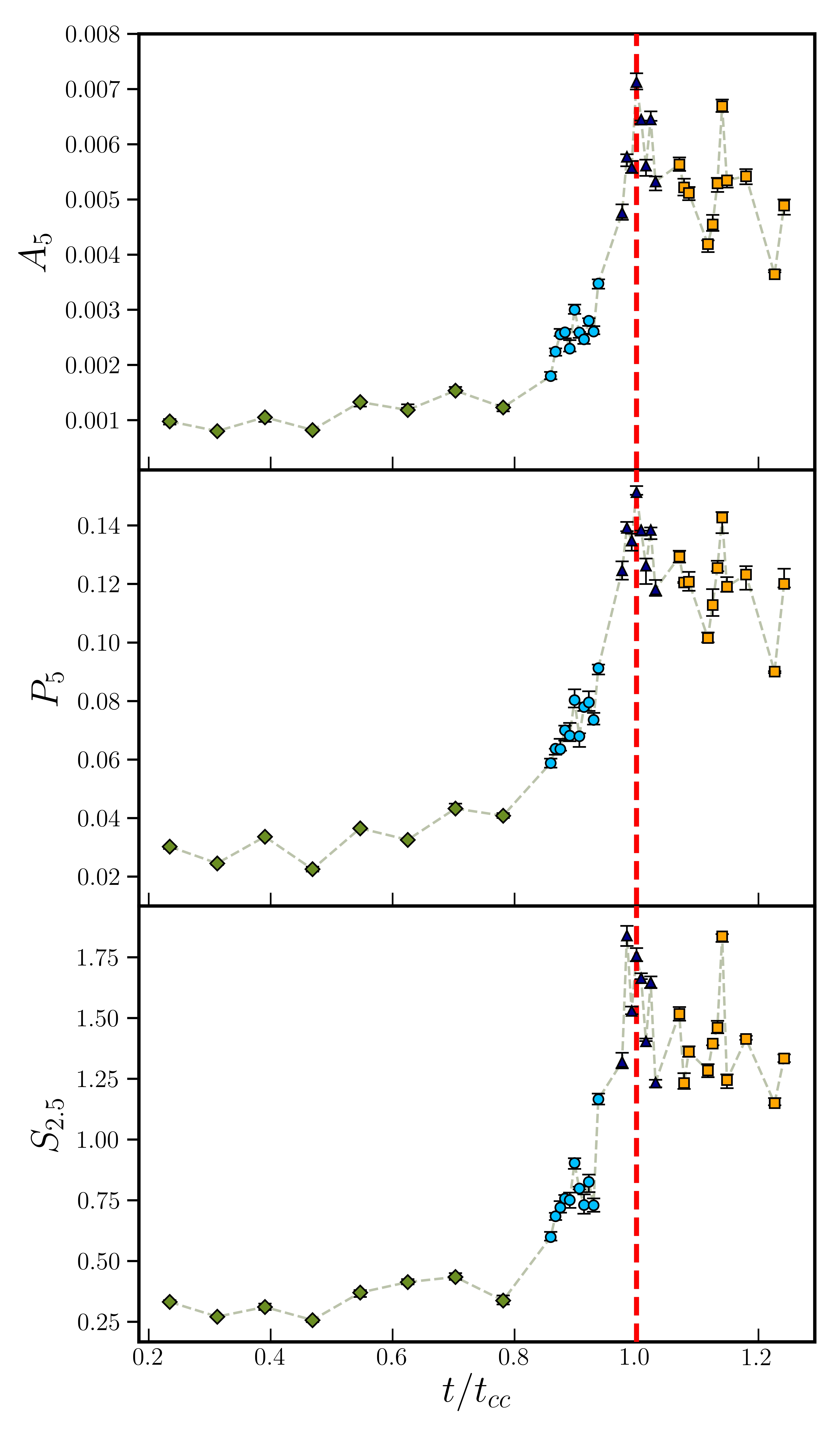} 
\caption{Time evolution of the nCRD parameters defined as in Figure
  \ref{fig:para} (see also Section \ref{sec:params}): $A_5$ (top
  panel), $P_5$ (central panel) and $S_{2.5}$ (bottom panel). Time is
  normalized to $t_{\rm CC}$, which is also marked by the vertical red
  dashed line.  The symbol shapes and colors are as in Figure
  \ref{fig:rc}. All the three parameters show a increasing trend with
  time, reaching the peak value at CC (blue triangles) and then
  remaining large and essentially constant, with some fluctuations,
  during the late, gravothermal oscillation stage (yellow squares).}
\label{fig:crdparams}
\end{figure}

Following the definitions above, we measured the three parameters for
all the snapshots under investigation.  To estimate their
uncertainties we took into account the dominant source of error,
namely the uncertainty of the half-mass radius as obtained from the
King fit to the density profile. To this end, for every snapshot we
re-determined the nCRD using all the stars with $V<V_{\rm cut}$
included within $R_n = 0.5 \times (r_h + \epsilon^+_h)$, where
$\epsilon^+_h$ is the upper error on $r_h$, and we measured the
corresponding values of the three parameters.  The difference between
this value and that obtained for $R_n=0.5 \times r_h$ is then adopted
as upper error on each parameter.  To estimate the lower uncertainty
we repeated the analogous procedure adopting $R_n = 0.5 \times (r_h
-\epsilon^-_h)$, $\epsilon^-_h$ being the lower error on $r_h$.
Figure \ref{fig:crdparams} shows the time evolution of $A_5$ (top
panel), $P_5$ (central panel), $S_{2.5}$ (bottom panel). The time is
normalized to $t_{\rm CC}$, which is also marked by the vertical red
dashed line, and the snapshots are plotted according to the color code
defined in Figure \ref{fig:lagrange_rad}.  As expected, all the
parameters show an increasing trend with time. In addition, the trend
is strikingly similar in the three cases: an almost constant behavior
is observed at the early evolutionary times (green points), then a
rapid increase occurs during the pre-CC stage (cyan points), up to the
achievement of maximum values at the CC epoch (blue points), followed
by the post-CC gravothermal oscillations stage during which the
parameters moderately fluctuate, but remain stable slightly below the
maximum reached at $t_{\rm CC}$ (yellow points).  To elaborate, $A_5$
remains constant around $A_5=0.001$ for most of the time ($t\lsim 0.8
t_{\rm CC}$), then it rapidly increases by a factor of 7 at CC, and it
slightly decreases and fluctuates around $A_5\sim 0.005$ later
on. Similarly, the $P_5$ parameter stays essentially constant around
$P_5\sim 0.03$ in the early epochs, then the increasing stellar
density in the cluster central regions increases it by a factor of 5
at CC, and finally the parameter stabilizes (with some fluctuations)
around $P_5\sim 0.12$ during the post-CC phase.  An analogous
evolutionary pattern is observed also for $S_{2.5}$: it is stable at
$S_{2.5}\sim 0.3$ at early times, then it shows a rapid increase, by a
factor of 6, reaching $S_{2.5}\sim 1.8$ at CC, and it further settles
around $S_{2.5}\sim 1.3$, with the usual fluctuations, during the
gravothermal oscillation stage.

\subsection{Sensitivity of the parameters to the assumptions}
\label{sec:sensitivity}
Of course, the exact shape of the nCRD (and, consequently, the values
of the newly defined parameters) depends on the assumptions used for
its construction, in particular the values of radial distance ($R_n$)
and the magnitude cut ($V_{\rm cut}$) adopted for the star sample
selection. In this section we thus explore the effects of modifying
these values.

\begin{figure}
\centering
\includegraphics[scale=0.5]{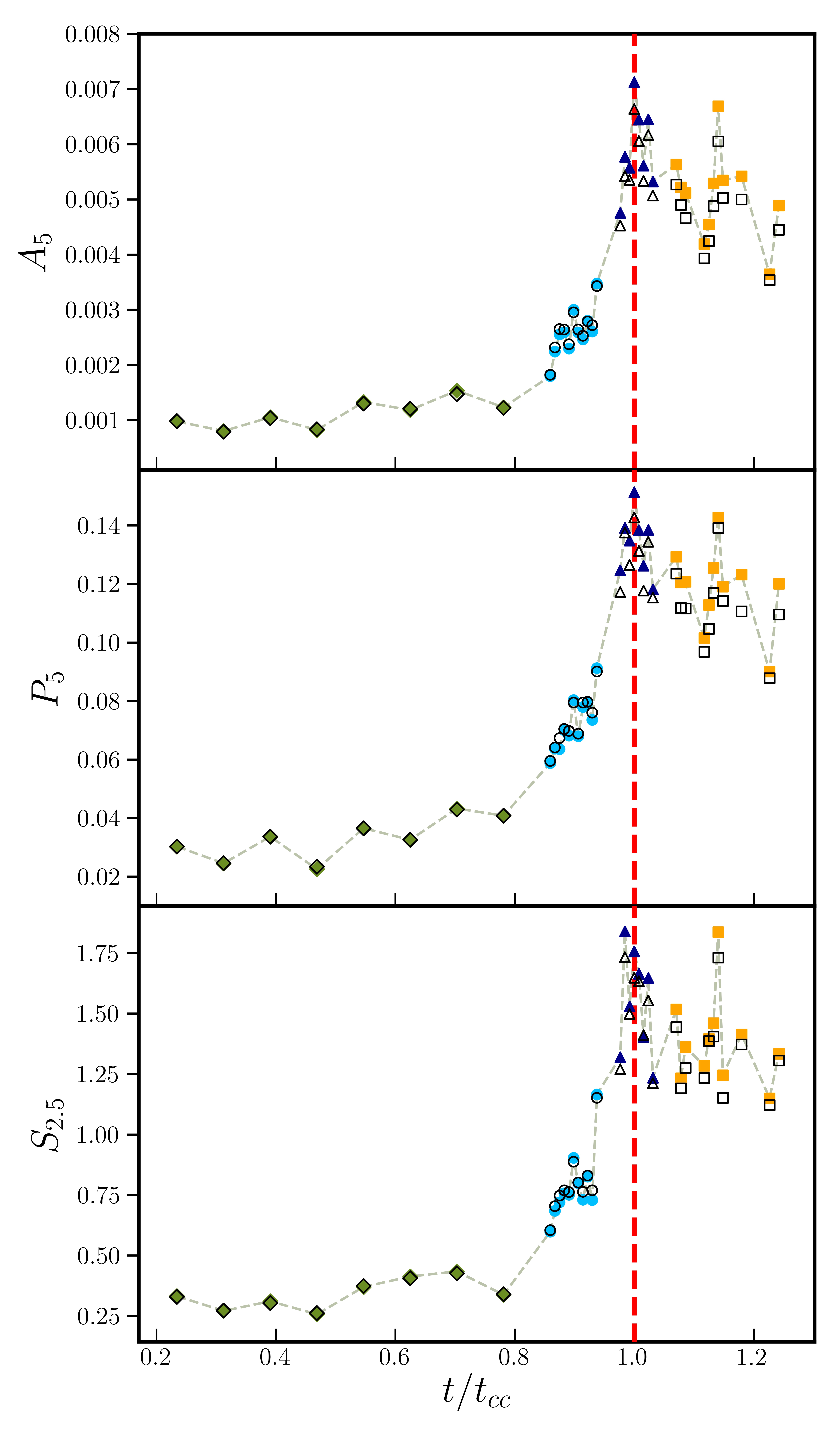} 
\caption{Comparison between the parameters obtained from nCDRs built
  using $R_n = 0.5 \times r_h$ (colored solid symbols, the same as in
  Fig. \ref{fig:crdparams}), and those obtained from nCDRs built using
  $R_n = 0.5 \times 65\arcsec$ (empty symbols).}
\label{fig:rn_effect}
\end{figure}

The (small) importance of varying $R_n$ can be already caught from the
size of the errors associated to each parameter (see error bars in
Fig. \ref{fig:crdparams}). However, we also investigated the effect of
a stronger variation, starting from the evidence that the King model
best-fitting a pre-CC snapshot (in particular, the one extracted at 10
Gyr) well reproduces also the density profile observed at later times
once the central cusp is excluded (see Fig. \ref{fig:cc}). We thus
assumed $R_n=0.5 \times 65\arcsec$ for all snapshots, this value being
half the cluster half-mass radius at 10 Gyr. The corresponding values
of the three parameters are shown as triangles in Figure
\ref{fig:rn_effect}. The comparison with the values obtained by
assuming the best-fit half-mass radius of each snapshot (circles, the
same as in Fig. \ref{fig:crdparams}) clearly confirms that the impact
of $R_n$ is negligible, at least for variations as large as $\sim
15\%$ of its value.

\begin{figure}
\centering
\includegraphics[scale=0.6]{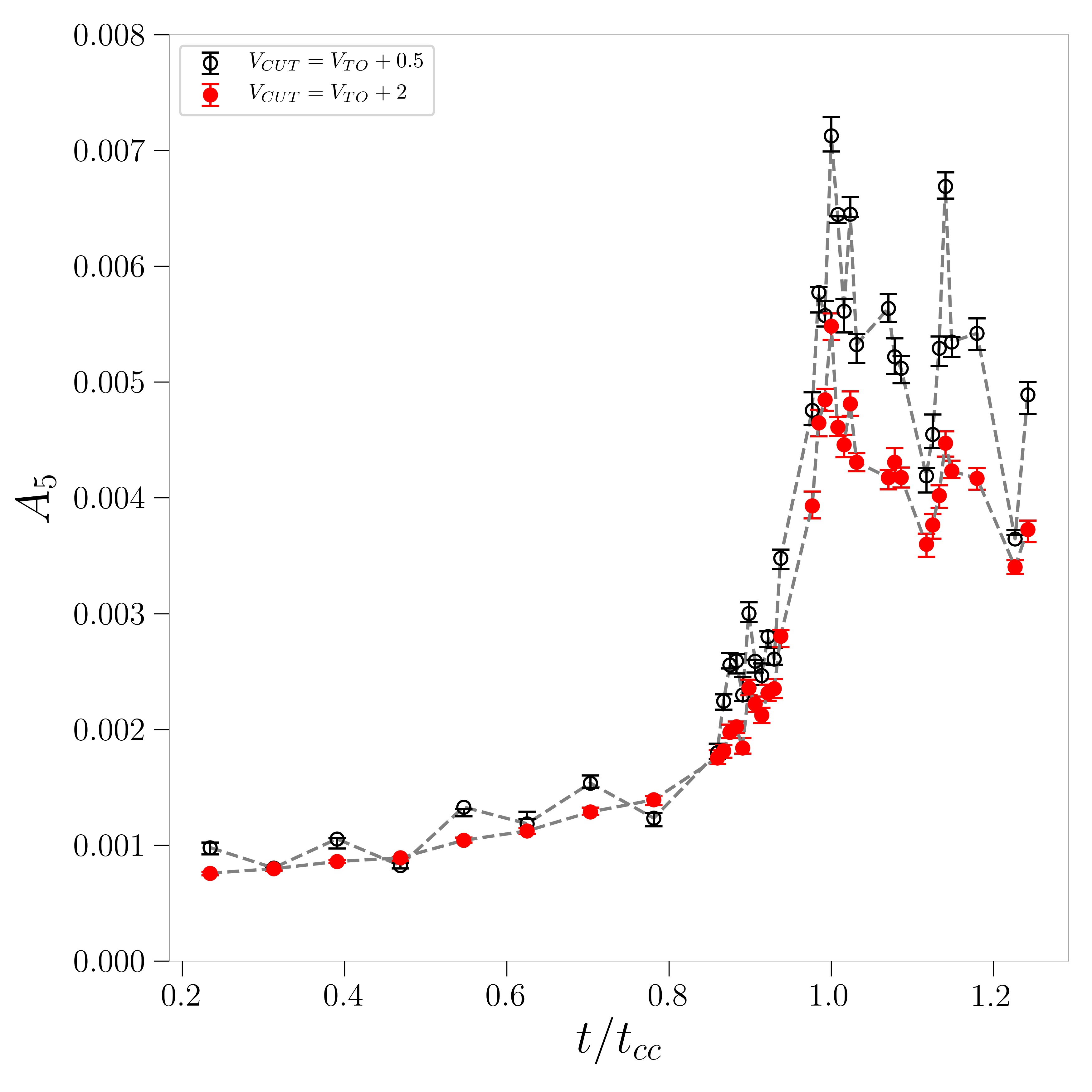}
\caption{Time evolution of the $A_5$ parameter measured from the nCRDs
  built with different magnitude cuts: $V_{\rm cut}=V_{\rm TO}+0.5$
  (empty black circles) and $V_{\rm cut}=V_{\rm TO}+2$ (solid red
  circles).}
\label{fig:area_vto0.5_2}
\end{figure}

As for the magnitude cut, the results shown so far have been obtained
by adopting a relatively bright threshold, just half magnitude below
the MS-TO: $V_{\rm cut}=V_{\rm TO} +0.5$.  This was done to allow the
observational measure of the nCRD parameters also in high-density
GGCs, where reaching deeper limits with a reasonable level of
completeness in the innermost regions of the system is still very
hard, even with HST data.  However, a fainter magnitude cut would
include a significantly larger sample of stars, thus offering the
advantage of a larger statistics. We thus explored the nCRDs obtained
with $V_{\rm cut}=V_{\rm TO} +2$ and measured from them the three
parameters defined above, to check whether they trace more or less
efficiently the cluster dynamical aging.  Figure
\ref{fig:area_vto0.5_2} shows the comparison between the time
evolution of $A_5$ as obtained for the two magnitude cuts: $V_{\rm
  cut}=V_{\rm TO}+ 0.5$ in black, a limit 1.5 magnitudes fainter in
red.  As can be seen, the overall trend is perfectly consistent in the
two cases, thus confirming that the increase of the $A_5$ parameter as
a function of time does not depend on the details of its own
definition, but traces, instead, the structural changes of the nCRD
due to the effects of dynamical evolution.  Indeed, the trend during
the early and pre-CC phases is virtually indistinguishable in the two
cases, except for a less noisy behavior for $V_{\rm cut}=V_{\rm TO}
+2$ due to the increased statistics.  Instead, the sensitivity of
$A_5$ during the CC and the post-CC phases appears systematically
reduced in the case of the deeper magnitude cut, with an increase of a
factor of $\sim 5$ (instead of $\sim 7$) with respect to the values
measured in the early snapshots.  This is consistent with the fact
that assuming a fainter magnitude threshold corresponds to including
stars of smaller masses in the analysis, which are less affected by
the dynamical evolutionary processes occurring in the cluster center.
Analogous dependencies on the adopted magnitude cut are also found for
the parameters $P_5$ and $S_{2.5}$.  Hence, as global result, we
conclude that the sensitivity of the nCRD parameters to the cluster
dynamical evolution tends to decrease with fainter magnitude cuts, and
$V_{\rm cut}=V_{TO}+0.5$, in spite of smaller numbers of stars, looks
as the best compromise between large enough statistics and good
efficiency to distinguish among different dynamical evolutionary
stages.

\section{Discussion and summary}
\label{sec:discussion}
In this paper we have presented the first results of a study aimed at
defining new empirical parameters that use the inner radial
distribution of cluster stars, to characterize the different dynamical
evolutionary phases experienced by dense stellar systems.  To this
end, we have used a Monte Carlo simulation following the `typical'
dynamical evolution of a GC, from an initial progressive contraction
of the core, to the CC event, and the gravothermal oscillations during
the post-CC phase.
  
A total of 38 time snapshots sampling these different stages have been
extracted from the simulation and analyzed by closely following the
steps usually taken in the analysis of an observational data set. We
used only projected (instead of three-dimensional) quantities and a
reasonable cut in magnitude to mimic the observational approach, where
photometric incompleteness and/or exposure times can severely limit
the extension in magnitude of the sample. In addition, the same
approach used in observational works has been applied to the numerical
data for determining the star density profile and its best-fit King
solution.  The analysis of the star density profiles extracted from
the simulation shows that the central cusp developing during CC is not
erased by the subsequent gravothermal oscillations; hence the central
cusp remains as a stable feature and characterizes the star density
profile also during the post-CC phase. However, a preliminary
inspection of the simulations shows that the slope of the central cup
can vary during the post-CC phase and its operational detectability
from observed data may present some difficulties.  A study aimed at a
detailed characterization of the key properties of the central density
cusp will be the specific subject of a future paper (Bhavana Bhat et
al., 2021, in preparation).

We then used the simulation to explore new ways of determining the
dynamical evolutionary stage of star clusters from the global
properties of their stellar population (instead of specific exotic
species such as, e.g., BSSs).  To this purpose we constructed and
analyzed the nCRD of each snapshot using all the stars brighter than
0.5 magnitude below the MS-TO and located within a projected distance
$R_n =0.5\times r_h$ from the center (see Section
\ref{sec:nCRD}). These showed an intriguing level of sensitivity to
the dynamical evolution of the cluster. Indeed, the shape of the nCRDs
varies significantly as a function of the cluster dynamical state and
allows a clear identification of the various fundamental stages of a
cluster evolution (the pre-CC, the CC, and the post-CC phases). We
have introduced three parameters (named $A_5$, $P_5$ and $S_{2.5}$)
that quantify the morphological changes of the nCRD as a function of
time and turned out to be effective diagnostics of the cluster
dynamical age. The three parameters show similar trends with time,
mirroring the host cluster dynamical evolution. After an early phase
(lasting $\sim 8$ Gyr in our simulation) in which they are essentially
constant, they rapidly increase reaching a maximum at the CC epoch. We
estimate that as the cluster approached CC, they grow by a factor of
5-7. The post-CC evolution yields to a slight decrease of the values
of all the parameters. However, in spite of some fluctuations, their
average value remains significantly larger than those typical of the
pre-CC phase. From an observational point of view this is one of the
most relevant aspect. Indeed, the fact that the values of the
parameters in the post-CC stages remain significantly larger than
those in the early phases offers the concrete possibility of clearly
distinguishing highly evolved GCs also in those cases where the
central density cusp detection is uncertain.

The results of this first exploratory work paves the way to a series
of future investigations in which we will broaden the range of initial
conditions and study their impact on the empirical parameters defined
here. By changing the initial structural properties of the simulated
clusters, we will compare the three parameters determined in stellar
systems that, after one Hubble time of evolution, have reached
different dynamical states. We will also include populations of
primordial binaries, which are known to halt the core contraction
earlier in the cluster evolution and at lower concentrations
\citep[see, e.g.,][]{vesperinichernoff94, trenti+07,
  chatterjee+10}. The main differences between the values of $A_5$,
$P_5$, and $S_{2.5}$ in simulations with and without primordial
binaries are expected during the advanced phases of the evolution,
towards CC and post-CC, when the milder contraction of the clusters
with primordial binaries might lead to a different and/or less extreme
evolution of these parameters. The study presented here will also be
further extended to explore the effects of different retention
fractions of dark remnants (neutron stars and black holes; see, e.g.,
\citealp{alessandrini+16, giersz+19, kremer+20, kremer+21, gieles+21},
for some studies on the dynamical effects of dark remnants).  Finally,
a forthcoming paper will be dedicated to build nCRDs and determine the
three parameters here defined in a sample of observed star
clusters. This requires photometric observations {\it(i)} with a high
enough angular resolution to resolve individual stars even in the
innermost cluster regions, {\it (ii)} sampling each system at least
out to $0.5 \times r_h$, and {\it (iii)} deep enough to reach a few
magnitudes below the MS-TO. These requirements are achieved by most
HST and adaptive-optics assisted observations currently available for
many GCs, thus making the determination of the three parameters from
observations relatively straightforward, although particular care is
needed to deal with typical observational difficulties as the
photometric incompleteness, differential reddening and Galactic field
contamination. We will discuss the relation between these parameters
and other dynamical indicators (in particular, the $A^+$ parameter
measured from BSSs; see the Introduction), thus providing quantitative
assessments of the operational ability of the three nCRD diagnostics
to distinguish dynamically-young GCs, from systems in advanced states
of dynamical evolution.


\vskip1truecm We thank the anonymous referee for comments and
suggestions that helped us to improve the paper. This work is part of
the project Cosmic-Lab at the Physics and Astronomy Department
``A. Righi'' of the Bologna University (http://www.cosmic-lab.eu/
Cosmic-Lab/Home.html). The research was funded by the MIUR throughout
the PRIN-2017 grant awarded to the project Light-on-Dark (PI:Ferraro)
through contract PRIN-2017K7REXT.

\appendix
\section{Simulations with different initial conditions}
\label{sec:appendix}
The parameter space of possible initial conditions for realistic
simulations of GCs is huge, including variations in the initial values
of $W_0$, scale radii, number of stars, primordial binary fraction,
dark remnant retention fraction. Hence, a series of forthcoming papers
will be devoted to accurately explore the effects that different
initial conditions can have on the values of the three proposed new
indicators and their time evolution.
  
\begin{figure}
\centering
\includegraphics[scale=0.6]{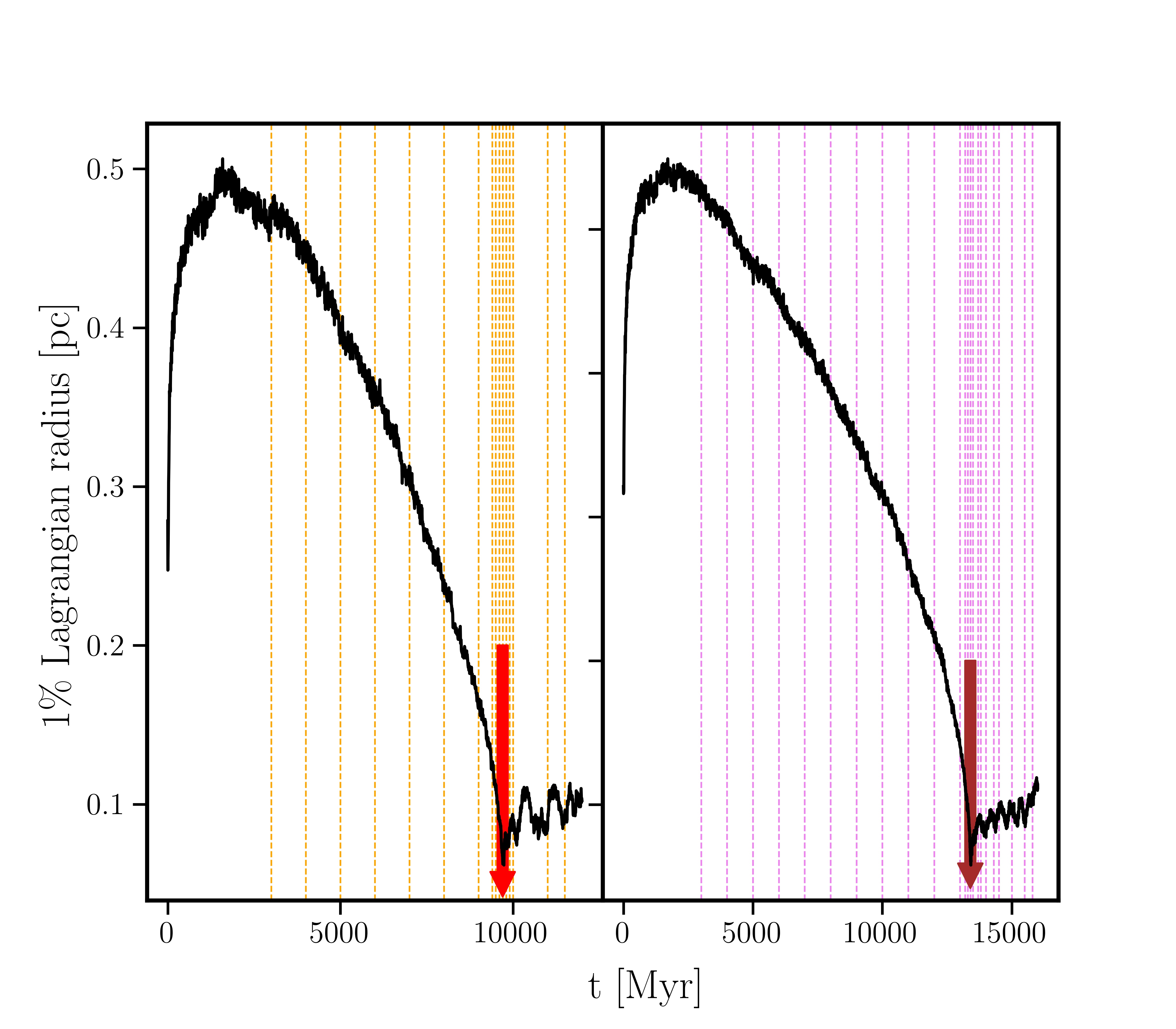} 
\caption{Time evolution of the 1\% Lagrangian radius (in pc) in 250K
  simulation (yellow, left panel) and the W05 run (magenta, right
  panel). The vertical lines correspond to the time snapshots for
  which we determined the values of $A_5$, $P_5$ and $S_{2.5}$ shown
  in Figure \ref{fig:param_appendix}.  The CC time is marked with a
  large red arrow: $t_{\rm CC}=9.7$ Gyr in the 250K run (left panel),
  $t_{\rm CC}=13.4$ Gyr in the W05 simulation (right panel).}
\label{fig:lagr_appendix}
\end{figure}

In this section we present just a first investigation of this issue,
by analyzing two additional simulations where only one initial
condition is varied at a time, with respect to the reference run
discussed in the main text (hereafter, REF run). In the first one
(hereafter, W05 run), we changed the value of the King dimensionless
potential ($W_0= 5$), while keeping the same initial number of stars
and half-mass radius, and assuming the same galactocentric distance as
in the reference simulation. In the second run (hereafter, 250K run),
we have followed the evolution of a system with half of the number of
particles (N=250K) and kept the same $W_0$, half-mass radius, and
galactocentric distance used for the REF simulation. The initial mass
in the W05 run is the same as in the REF model, while it is $\sim 1.6
\times 10^5 M_\odot$ initially in the 250K simulation (and $\sim 5.7
\times 10^4 M_\odot$ at 12 Gyr).  Figure \ref{fig:lagr_appendix} shows
the time evolution of the 1\% Lagrangian radius of these two
simulations. The overall trend is very similar to that shown in Figure
\ref{fig:lagrange_rad} for the REF run. However, CC occurs earlier (at
$t_{\rm CC} = 9.7$ Gyr) in the case of the less massive cluster (250K
run; left panel), while it is delayed by almost 1 Gyr for the
initially less concentrated cluster (W05 run, where $t_{\rm CC} =
13.4$ Gyr; right panel). The vertical lines mark the time snapshots
extracted from these simulations, which have been analyzed following
the same procedures and adopting the same assumptions discussed in the
main text for the REF simulation.

\begin{figure}
\centering
\includegraphics[scale=0.5]{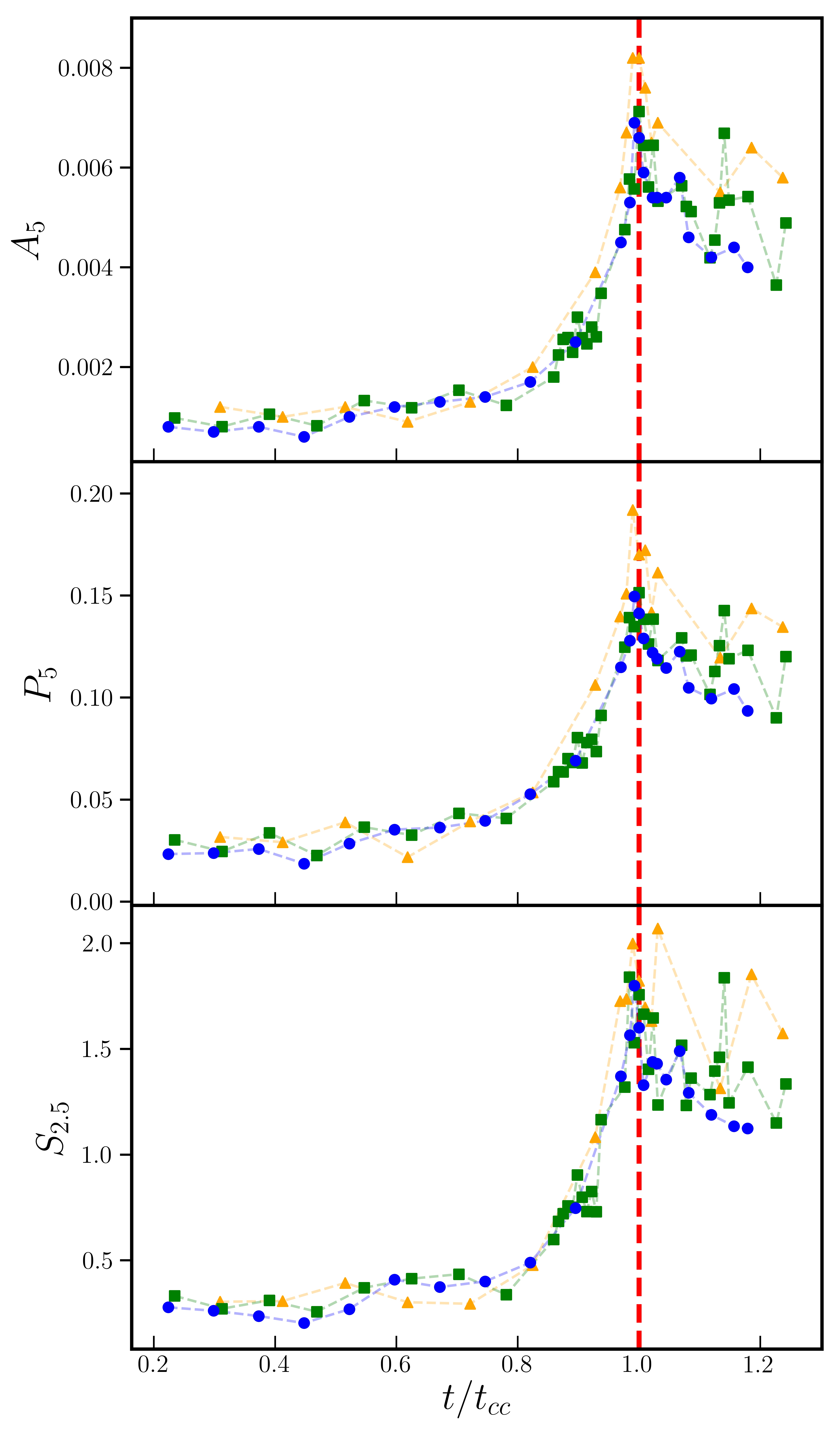}
\caption{Time evolution of the nCRD parameters in simulations 250K
  (yellow triangles), W05 (blue circles), and REF (green squares, the
  same points as in Figure \ref{fig:crdparams}). From top to bottom,
  the three panels refer to parameters $A_5$, $P_5$ and $S_{2.5}$.
  Time is normalized to each respective value of $t_{\rm CC}$.}
\label{fig:param_appendix}
\end{figure}

The resulting time dependence of the three parameters is shown in
Figure \ref{fig:param_appendix}, where the yellow circles correspond
to the 250K simulation, the blue circles refer to run W05 and, for the
sake of comparison, we overplotted also the results of the REF
simulation in green (same points as in Figure
\ref{fig:crdparams}). Along the x-axis, the time is normalized to the
respective values of $t_{\rm CC}$. The comparison shows that the
differences among the adopted initial conditions have a negligible
effect on both the absolute values and the time dependence of the
three parameters, thus further reinforcing the conclusion that $A_5$,
$P_5$, and $S_{2.5}$ are powerful indicators of GC internal dynamical
evolution.

\bibliography{dynam_simu}{}
\bibliographystyle{aasjournal}



\end{document}